# Dephasing of Transverse Spin Current in Ferrimagnetic Alloys

Youngmin Lim[1], Behrouz Khodadadi[1], Jie-Fang Li[2], Dwight Viehland[2], Aurelien Manchon[3,4], Satoru Emori[1,*]

1. Department of Physics, Virginia Tech, Blacksburg, VA 24061, USA
2. Department of Materials Science and Engineering, Virginia Tech, Blacksburg, VA 24061, USA
3. Physical Science and Engineering Division (PSE), King Abdullah University of Science and Technology (KAUST), Thuwal 23955-6900, Saudi Arabia
4. Aix-Marseille Univ, CNRS, CINaM, Marseille, France

* email: semori@vt.edu

**It has been predicted that transverse spin current can propagate coherently (without dephasing) over a long distance in antiferromagnetically ordered metals. Here, we estimate the dephasing length of transverse spin current in ferrimagnetic CoGd alloys by spin pumping measurements across the compensation point. A modified drift-diffusion model, which accounts for spin-current transmission through the ferrimagnet, reveals that the dephasing length is about 4-5 times longer in nearly compensated CoGd than in ferromagnetic metals. This finding suggests that antiferromagnetic order can mitigate spin dephasing – in a manner analogous to spin echo rephasing for nuclear and qubit spin systems – even in structurally disordered alloys at room temperature. We also find evidence that transverse spin current interacts more strongly with the Co sublattice than the Gd sublattice. Our results provide fundamental insights into the interplay between spin current and antiferromagnetic order, which are crucial for engineering spin torque effects in ferrimagnetic and antiferromagnetic metals.**

**I. INTRODUCTION**

A spin current is said to be coherent when the spin polarization of its carriers, such as electrons, is locked in a uniform orientation or precessional phase. How far a spin current propagates before decohering underpins various phenomena in solids [1,2]. Spin decoherence can generally arise from *spin-flip scattering*, where the carrier spin polarization is randomized via momentum scattering [3,4]. In magnetic materials, electronic spin current polarized transverse to the magnetization can also decohere by *dephasing*, where the total carrier spin polarization vanishes due to the destructive interferences of precessing spins (i.e., upon averaging over the Fermi surface) [5–9]. In typical ferromagnetic metals (FMs), the dephasing length $\lambda_{dp}$ is only ≈1 nm [5–7,10] whereas the spin-flip (diffusion) length $\lambda_{sf}$ may be considerably longer (e.g., ≈10 nm) [3,4], such that dephasing dominates the decoherence of transverse spin current.

Figure 1(a) qualitatively illustrates the dephasing of a coherent electronic spin current in a FM spin sink. In this particular illustration, a coherent ac transverse spin current is excited by ferromagnetic resonance (FMR) spin pumping [11], although in general a coherent transverse spin current may be generated by other means (e.g., dc electric current spin-polarized transverse to the magnetization [5,6,12]). This spin current, carried by electrons, then propagates coherently through the normal metal spacer (e.g., Cu, where $\lambda_{sf}$~100 nm is much greater than the typical spacer thickness) [13]. However, this spin current enters the FM spin sink with a wide distribution of incident wavevectors[1], spanned by the Fermi surface of the FM. Electronic spins with different wavevectors require different times to reach a certain depth in the FM, thereby spending different times in the exchange field. Thus, even though these electronic spins enter the FM with the same phase, they precess about the exchange field in the FM by different amounts. Within a few atomic monolayers in the FM, the transverse spin polarization averages to zero; the spin current dephases within a short length scale $\lambda_{dp}$ ≈ 1 nm.

---

[1] An insulating tunnel barrier is known to filter the incident wavevectors to a narrow distribution [94]. This filtering effect can reduce dephasing and thus extend $\lambda_{dp}$.

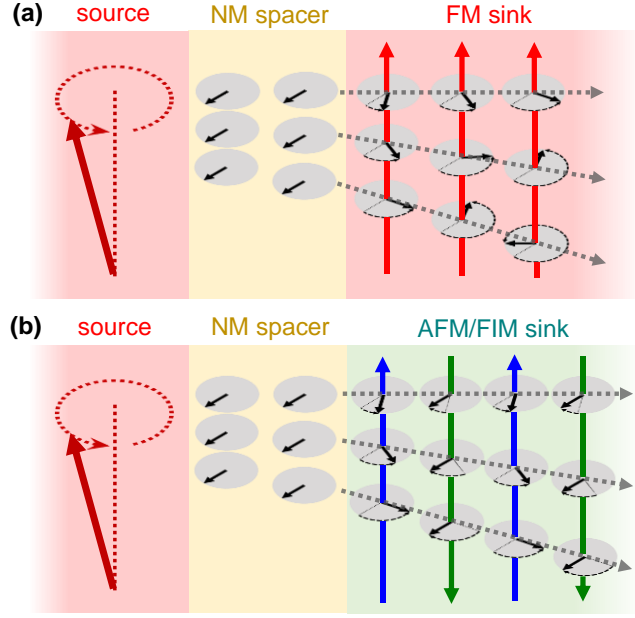

**FIG. 1.** Dephasing of a coherent transverse spin current excited by ferromagnetic resonance (FMR) in the spin source. The spin current carried by electrons is coherent in the normal metal (NM) spacer layer (indicated by the aligned black arrows), but enters the spin sink with different incident wavevectors (dashed gray lines). (a) In the ferromagnetic metal (FM) spin sink, the propagating spins accumulate different precessional phases in the ferromagnetic exchange field (red vertical arrows) and completely dephase within a short distance. (b) In the ideal antiferromagnetic metal (AFM) or ferrimagnetic metal (FIM), the spin current does not dephase completely in the alternating antiferromagnetic exchange field (blue and green vertical arrows), as any precession at one sublattice is compensated by the opposite precession at the other sublattice. In the case of a FIM that is an alloy of a transition metal (TM, such as Co) and rare-earth metal (RE, such as Gd), the TM constitutes one sublattice (e.g., blue arrows) and the RE constitutes the other sublattice (e.g., green arrows).

Transverse spin currents in antiferromagnetically ordered metals have been predicted to exhibit longer $\lambda_{dp}$ [14–17]. This prediction may apply not only to intrinsic antiferromagnetic metals (AFMs) but also compensated ferrimagnetic metals (FIMs), which consist of transition-metal (TM) and rare-earth-metal (RE) magnetic sublattices that are antiferromagnetically coupled to each other [18]. In the ideal case as illustrated in Fig. 1(b), the spin current interacts with the staggered antiferromagnetic exchange field whose direction alternates at the atomic length scale. The propagating spins precess in alternating directions as they move from one magnetic sublattice to the next, such that spin dephasing is suppressed over multiple monolayers. This cancellation of dephasing in AFMs and FIMs is analogous to spin rephasing by π-pulses (Hahn spin echo method) in nuclear magnetic resonance [19], which has recently inspired several approaches of mitigating decoherence of qubit spin systems [20–22].

The above idealized picture for extended coherence in antiferromagnetically ordered metals (Fig. 1(b)) assumes a spin current without any scattering and simple layer-by-layer alternating collinear magnetic order. Finite scattering, spin-orbit coupling, and complex magnetization states in real materials may disrupt transverse spin coherence [23–25]. The transverse spin coherence length $\lambda_c$ accounting for both spin-flip scattering and spin dephasing is given by [10,26],

$$\frac{1}{\lambda_c} = \mathrm{Re}\left[\sqrt{\frac{1}{\lambda_{sf}^2} - \frac{i}{\lambda_{dp}^2}}\right]. \qquad (1)$$

Thus, in real AFMs and FIMs, a shorter coherence length results from reduced $\lambda_{sf}$ due to increased spin-flip rates, or reduced $\lambda_{dp}$ due to momentum scattering and non-collinear magnetic order that prevents perfect cancellation of dephasing [23,24]. Most experiments on AFMs (e.g., polycrystalline IrMn) indeed show short coherence lengths of ≈1 nm [7,27–30].

Nevertheless, a recent experimental study utilizing a spin-galvanic detection method [31–35] has reported a long coherence length of >10 nm at room temperature in FIM CoTb [18]. The report in Ref. [18] is quite surprising considering the strong spin-orbit coupling of CoTb, primarily from RE Tb with a large orbital angular momentum, which can result in increased spin-flip scattering [36–38] and noncollinear sperimagnetic order [39–41]. TM and RE elements also tend to form amorphous alloys [39–42], whose structural disorder may result in further scattering and deviation from layer-by-layer antiferromagnetic order. It therefore remains a critical issue to confirm whether the cancellation of dephasing (as depicted in Fig. 1(b)) actually extends transverse spin coherence in antiferromagnetically ordered metals, particularly structurally disordered FIMs.

Here, we test for the suppressed dephasing of transverse spin current in ferrimagnetic alloys. Our experimental test consists of FMR spin pumping measurements [7,28] on a series of amorphous FIM CoGd spin sinks, which exhibit significantly weaker spin-orbit coupling than CoTb due to the nominally zero orbital angular momentum of RE Gd. Our experimental results combined with a modified drift-diffusion model [9,10,43,44] reveal that spin dephasing is indeed partially cancelled in nearly compensated CoGd, with $\lambda_{dp}$ extended by a factor of 4-5 compared to that for FMs. Moreover, we find evidence that transverse spin current interacts more strongly with the TM Co sublattice than the RE Gd sublattice. Our results suggest that, even in the presence of substantial structural disorder, the antiferromagnetically coupled sublattices in FIMs can mitigate the decoherence of transverse spin current. On the other hand, the maximum $\lambda_{dp}$ of ≈5 nm in FIM CoGd found here is quantitatively at odds with the report of $\lambda_{dp} > 10$ nm in FIM CoTb [18]. Our study sheds light on the interaction of transverse spin current with antiferromagnetic order, which underpins spin torque control of FIMs and AFMs for fast spintronic devices [45–47].

## II. FILM GROWTH AND STATIC MAGNETIC PROPERTIES

We deposited spin-valve-like stacks of Ti(3)/Cu(3)/Ni$_{80}$Fe$_{20}$(7)/Cu(4)/Co$_{100-x}$Gd$_x$(*d*)/Ti(3) (unit: nm) with *x* = 0, 20, 22, 23, 25, 28, and 30 by dc magnetron sputtering on Si/SiO$_2$ substrates. The base pressure in the deposition chamber was better than $8 \times 10^{-8}$ Torr. The Ar sputtering gas pressure was 3 mTorr. The Ti(3)/Cu(3) seed layer promotes the growth of Ni$_{80}$Fe$_{20}$ with low Gilbert damping and minimal inhomogeneous linewidth broadening, whereas the Ti(3) capping layer protects the stack from oxidation. FIM Co$_{100-x}$Gd$_x$ films with various Gd concentrations (x in atomic %) were deposited by co-sputtering Co and Gd targets at different Gd sputtering powers (resulting in an uncertainty in composition of ≈±0.5 at.% Gd), except for Co$_{80}$Gd$_{20}$ and Co$_{70}$Gd$_{30}$ films that were deposited by sputtering compositional alloy targets. The deposition rate of each target was calibrated by x-ray reflectivity and was set at 0.020 nm/s for Ti, 0.144 nm/s for Cu, 0.054 nm/s for NiFe, 0.011 or 0.020 nm/s for Co, 0.008 – 0.020 nm/s for Gd, 0.014 nm/s for Co$_{80}$Gd$_{20}$, and 0.012 nm/s for Co$_{70}$Gd$_{30}$.

We performed vibrating sample magnetometry (with a Microsense EZ9 VSM) to identify the magnetic compensation composition at room temperature for Co$_{100-x}$Gd$_x$ films. We measured Co$_{100-x}$Gd$_x$ as single-layer films and as part of the spin-valve-like stacks (NiFe/Cu/CoGd), both seeded by Ti(3)/Cu(3) and capped by Ti(3). CoGd in both types of samples exhibited identical results within experimental uncertainty. In the spin-valve-like stacks, the Cu spacer layer suppresses static exchange coupling between the NiFe and CoGd layers. This interlayer exchange decoupling is corroborated by magnetometry results (e.g., Fig. 2(a)) that indicate separate switching for the NiFe and CoGd layers.

Figure 2(b-g) summarizes the composition dependence of the static magnetic properties of Co$_{100-x}$Gd$_x$. Our results corroborate the well-known trend in ferrimagnetic alloys: the saturation magnetization converges toward zero and coercivity diverges near the composition at which the magnetic moments of the Co and Gd sublattices compensate. Comparing the results for different CoGd thicknesses, we find that the magnetization compensation composition shifts toward higher Gd content with decreasing CoGd thickness. A similar thickness dependence of the compensation composition in TM-RE FIM films has been seen in prior studies [18]. Since precise determination of the magnetic compensation composition was difficult for smaller $Co_{100-x}Gd_x$ thicknesses, we tentatively identify the magnetic compensation composition window to be $x \approx$ 22-25, which is what we find for 5-nm-thick CoGd. We also remark that the angular momentum compensation composition is expected to be ≈1 Gd at. % below the magnetic compensation composition, considering the *g*-factors of Co and Gd ($g_{Co} \approx$ 2.15, $g_{Gd}$ = 2.0) [48]. CoGd layers in our stack structures do not show perpendicular magnetic anisotropy [18,49–58], i.e., CoGd films here are in-plane magnetized [59–62].

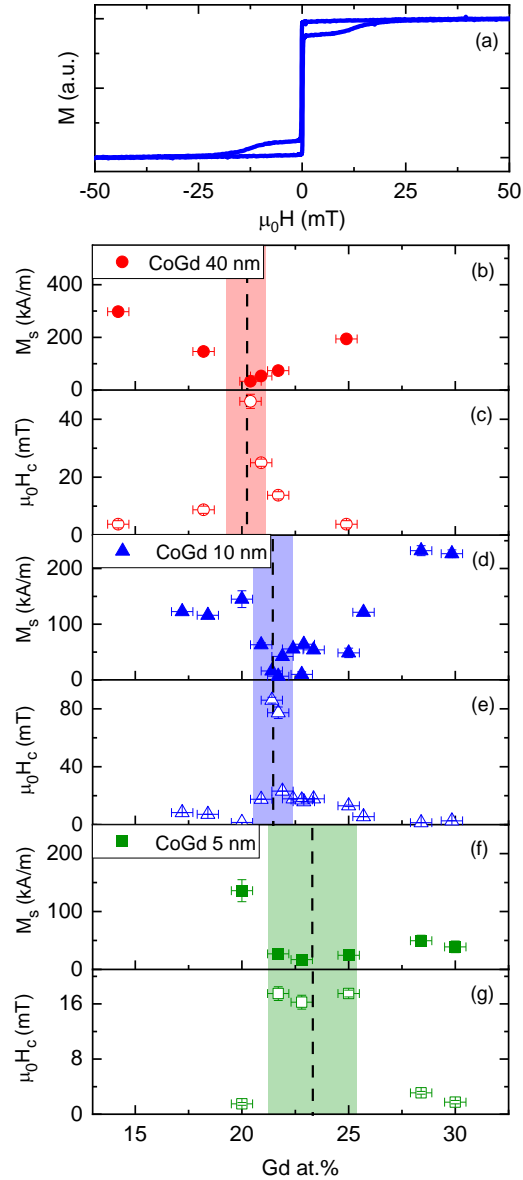

**FIG. 2.** (a) Hysteresis loop of $Ni_{80}Fe_{20}(7)/Cu(4)/Co_{75}Gd_{25}(13)$ (unit: nm). The $Ni_{80}Fe_{20}$ and $Co_{75}Gd_{25}$ magnetizations switch separately at in-plane fields ≈0.2 mT and ≈13 mT, respectively. (b,d,f) Saturation

magnetization $M_s$ and (c,e,g) coercivity $\mu_0H_c$ of $Co_{100-x}Gd_x$ with thicknesses of 40 nm (b,c), 10 nm (d,e), and 5 nm (f,g).

## III. CHARACTERIZATION OF SPIN TRANSPORT BY SPIN PUMPING

### A. FMR Spin Pumping Experiment

The multilayer stacks (described in Sec. II) in our study consist of a NiFe spin source and a $Co_{100-x}Gd_x$ spin sink, separated by a diamagnetic Cu spacer. A coherent spin current generated by FMR [11,13] in NiFe propagates through the Cu spacer and decoheres in the $Co_{100-x}Gd_x$ spin sink, yielding nonlocal Gilbert damping [7,28]. The Cu spacer layer suppresses exchange coupling – and hence direct magnon coupling – between the NiFe and CoGd layers [63]. The diamagnetic Cu spacer also accommodates spin transport mediated solely by conduction electrons, such that direct interlayer magnon coupling [17,64–66] does not play a role here[2].

In our FMR spin pumping experiments performed at room temperature, the half-width-at-half-maximum linewidth $\Delta H$ of the NiFe spin source is measured at microwave frequencies $f$ = 2-20 GHz. The details of the FMR measurement method are in Appendix A. The FMR response of the NiFe layer is readily separated from that of pure Co (x = 0), and CoGd did not yield FMR signals above our instrumental background (see Fig. 10 in Appendix A). Thus, as shown in Fig. 3, the Gilbert damping parameter $α$ for the NiFe layer is quantified from the $f$ dependence of $\Delta H$ through the linear fit,

$$\mu_0 \Delta H = \mu_0 \Delta H_0 + \frac{h}{g\mu_B} \alpha f, \qquad (2)$$

where $g \approx 2.1$ is the Landé $g$-factor of $Ni_{80}Fe_{20}$, $\mu_0$ is the permeability of free space, $h$ is Planck's constant, $\mu_B$ is the Bohr magneton, $\mu_0 \Delta H_0$ (< 0.2 mT) is the zero-frequency linewidth attributed to magnetic inhomogeneity [67]. For NiFe without a spin sink, we obtain $α_{\text{no-sink}} ≈$ 0.0067, similar to typically reported values for $Ni_{80}Fe_{20}$ [68,69].

A finite thickness $d$ of spin sink results in a damping parameter $α_{\text{w/sink}}$ that is greater than $α_{\text{no-sink}}$. For example, the damping increases significantly with just $d = 1$ nm of Co (Fig. 3(a)), suggesting substantial spin absorption by the spin sink. By contrast, a stack structure that includes an insulating layer of Ti-oxide before the spin sink does not show the enhanced damping (Fig. 3(a)). This observation is consistent with the Ti-oxide layer blocking the spin current [70,71] between the spin source and spin sink layers. Thus, the enhanced damping $\Delta α$ =

---

[2] We note that some of the electronic spin current might be converted into a magnonic spin current [95] at the Cu/CoGd interface. However, for the sake of simplicity, we assume the dominance of electronic spin transport in CoGd here.

$α_{w/sink} − α_{no-sink}$ is nonlocal in origin, i.e., due to the spin current propagating through the Cu spacer and decohering in the magnetic spin sink [7,10,11,13,28,43]. The decoherence of transverse spin current in the spin sink is then directly related to Δα. This spin pumping method based on nonlocal damping enhancement provides an alternative to the spin-galvanic method [18] that is known to contain parasitic voltage signals unrelated to spin transport [31–35], as discussed further in Sec. IV-C.

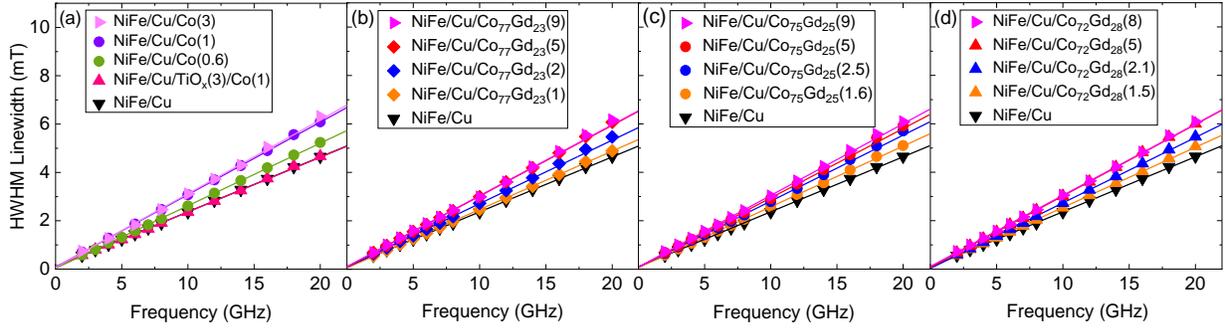

**FIG. 3.** (a) Half-width-at-half-maximum (HWHM) FMR linewidth versus frequency for stacks with a spin sink (NiFe/Cu/Co($d$: nm)), a stack without a spin sink (NiFe/Cu), and a stack with an insulating Ti-oxide spin blocker before the spin sink (NiFe/Cu/TiO$_x$/Co). (b-d) FMR linewidth versus frequency for stacks with different thicknesses ($d$: nm) of Co$_{100-x}$Gd$_x$ spin sinks, where x = 23 (b), 25 (c), and 28 (d). The slope (proportional to the damping parameter $α$, see Eq. (2)) saturates at $d ≈ 1$ nm for the FM Co spin sink (a), whereas the slope saturates at a much larger $d$ for the FIM Co$_{100-x}$Gd$_x$ spin sinks (b-d).

In contrast to the large Δα with an ultrathin FM Co spin sink, the damping enhancement with $d$ is more gradual for FIM CoGd sinks. Figure 3 shows exemplary linewidth versus frequency results for the spin sink compositions of Co$_{77}$Gd$_{23}$ (Fig. 3(b)), Co$_{75}$Gd$_{25}$ (Fig. 3(c)), and Co$_{72}$Gd$_{28}$ (Fig. 3(d)). A damping enhancement similar in magnitude to that of the 1-nm-thick FM Co spin sink is reached only when the CoGd thickness is several nm. This suggests that transverse spin-current decoherence takes place over a greater length scale in FIM spin sinks than in FM spin sinks.

In Fig. 4, we summarize our experimental results of transverse spin decoherence (i.e., Δα) as a function of spin sink thickness $d$. It can be seen that Δα for each spin sink composition saturates above a sufficiently large $d$. This apparent saturation thickness – related to how far the transverse spin current remains coherent [7,10] – changes markedly with the spin sink composition. With FM Co as the spin sink, the saturation of Δα occurs at $d ≈ 1$ nm, in agreement with $λ_{dp}$ reported before for FMs [7,10]. By contrast, Δα saturates at $d \gg 1$ nm for FIM CoGd sinks. This observation again implies that transverse spin current remains coherent deeper within FIM sinks than within FM sinks.

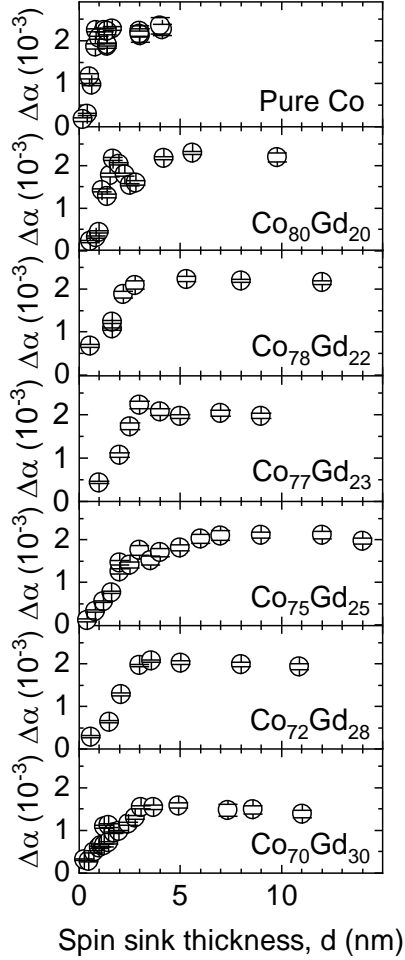

**FIG. 4.** Nonlocal damping enhancement Δ$\alpha$ versus spin sink thickness *d* for CoGd spin sinks with different compositions.

We now consider possible mechanisms of the longer decoherence lengths in FIM sinks. Increasing the Gd concentration dilutes the magnitude of the exchange field in $Co_{100-x}Gd_x$ alloys, as evidenced by the reduction of the Curie temperature from ≈1400 K for pure Co to ≈700 K for $Co_{70}Gd_{30}$ [72]. The diluted exchange field would lead to slower spin dephasing (i.e., longer $\lambda_{dp}$), thereby requiring a thicker spin sink for complete spin-current decoherence. This mechanism would yield $\lambda_{dp}$ that is inversely proportional to the Curie temperature [73], such that the spin-sink thickness $d$ at which Δ$\alpha$ saturates would increase monotonically with Gd content. However, we do not observe such a monotonic trend in Fig. 4. Rather, the saturation thickness appears to plateau or peak at a Gd content of x ≈ 25, which is close to the magnetic compensation composition window.

We therefore consider an alternative mechanism, where the antiferromagnetically coupled Co and Gd sublattices in the FIM alloys mitigate dephasing, as qualitatively illustrated in Fig. 1(b). This mechanism would be expected to maximize $\lambda_{dp}$ near the compensation composition. In the following subsection, we describe and apply a modified spin drift-diffusion model to estimate $\lambda_{dp}$ and examine the possible role of the antiferromagnetic order in the mitigation of dephasing in FIM CoGd.

**B. Modified Drift-Diffusion Model**

We wish to model our experimental results and quantify $\lambda_{dp}$ for the series of CoGd sinks. The conventional drift-diffusion model captures spin-flip scattering in nonmagnetic metals [11,74,75], but not spin dephasing that is expected to be significant in magnetic metals. This conventional model also predicts a monotonic increase of Δα with $d$ [11,74,75], whereas we observe in Fig. 4 a non-monotonic behavior where Δα overshoots before approaching saturation for some compositions of CoGd. For spin pumping studies with magnetic spin sinks, typical models assume that the transverse spin current decoheres by dephasing as soon as it enters the FM spin sink, i.e., $\lambda_{dp} = 0$ [13,75,76]. Others fit a linear increase of Δα with $d$ up to apparent saturation, deriving $\lambda_{dp} = 1.2 \pm 0.1$ nm for FMs [7,28,73]. However, it is questionable that this linear cut-off model applies in a physically meaningful way to our experimental results (Fig. 4), in which the increase of Δα to saturation is not generally linear.

We therefore apply an alternative model that captures the dephasing (i.e., precession and decay) of transmitted transverse spin current in the magnetic spin sinks by invoking the *transmitted spin-mixing conductance* $g_t^{\uparrow\downarrow}$ [5,9,10,43,44,77,78]. As illustrated in Fig. 5, $g_t^{\uparrow\downarrow}$ accounts for the spin current transmitted through the spin sink, in contrast to the conventional *reflected* spin-mixing conductance that accounts for the spin current reflected at the spin sink interface [7,79]. $g_t^{\uparrow\downarrow}$ is a function of the magnetic spin sink thickness $d$ that must vanishes in the limit of $d \gg \lambda_{dp}$, i.e., when the transverse spin current completely dephases in a sufficiently thick spin sink [5,9,77,78]. In conventional FM spin sinks, it is often assumed that $g_t^{\uparrow\downarrow} = 0$, which is equivalent to assuming $\lambda_{dp} = 0$ [79].

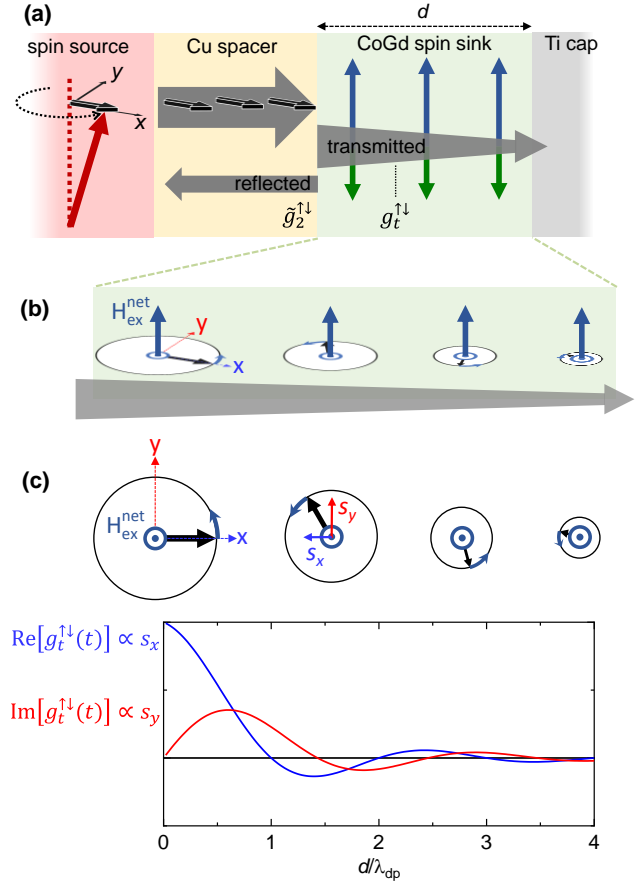

**FIG. 5.** (a) Cartoon schematic of transverse spin current transport in the spin-source/spacer/spin-sink stack. The FMR-driven spin source pumps a coherent transverse spin current (polarized along the *x*-axis) through the Cu spacer. The spin current reflected at the Cu (spacer)/CoGd (spin sink) interface is parameterized by the reflected spin-mixing conductance $\tilde{g}_2^{\uparrow\downarrow}$, whereas the spin current transmitted through the CoGd spin sink is parameterized by the transmitted spin-mixing conductance $g_t^{\uparrow\downarrow}$. (b) Illustration of the dephasing of the transverse spin current propagating in the CoGd spin sink. The shrinking circles represent the decay of the transverse spin polarization due to dephasing while it precesses about the effective net exchange field $H_{ex}^{net}$. (c) Illustration of the oscillatory decay of the transverse spin current. The complex transmitted spin-mixing conductance $g_t^{\uparrow\downarrow}$ captures the transmitted transverse spin polarization components $s_x$ and $s_y$.

Furthermore, $g_t^{\uparrow\downarrow}(d)$ is a complex value where the real and imaginary parts are comparable in magnitude [78]. $\text{Re}[g_t^{\uparrow\downarrow}(d)]$ and $\text{Im}[g_t^{\uparrow\downarrow}(d)]$ are related to the two orthogonal components of the spin current transmitted through the magnetic spin sink [9]. As illustrated in Fig. 5(a,b), the incident spin polarization is along the *x*-axis, whereas the *y*-axis is normal to both the incident spin polarization and the magnetic order of the spin sink. $\text{Re}[g_t^{\uparrow\downarrow}(d)]$ and $\text{Im}[g_t^{\uparrow\downarrow}(d)]$ represent the *x*- and *y*-components, respectively, of the transverse spin polarization (Fig. 5); these components oscillate and decay while the spin current dephases.

We approximate $g_t^{\uparrow\downarrow}(d)$ with an oscillatory decay function [9],

$$g_t^{\uparrow\downarrow}(d) = g_{t,0}^{\uparrow\downarrow} \left( \frac{\lambda_{dp}}{\pi d} \sin \frac{\pi d}{\lambda_{dp}} \pm i \left[ \left( \frac{\lambda_{dp}}{\pi d} \right)^2 \sin \frac{\pi d}{\lambda_{dp}} - \frac{\lambda_{dp}}{\pi d} \cos \frac{\pi d}{\lambda_{dp}} \right] \right) \exp\left(-\frac{d}{\lambda_{sf}}\right), \quad (3)$$

where $g_{t,0}^{\uparrow\downarrow}$ represents the interfacial contribution to $g_t^{\uparrow\downarrow}$. Equation (3) is identical to the function proposed by Kim [9], except that we incorporate an exponential decay factor (with $\lambda_{sf}$ = 10 nm as explained in Appendix B) that approximates incoherent spin scattering as an additional source of spin-current decoherence. The sign between the real and imaginary terms in Eq. (3) represents the net precession direction of the transverse spin polarization: the positive sign indicates precession about an exchange field *along* the net magnetization, whereas the negative sign indicates precession about an exchange field *opposing* the net magnetization. Although the effective exchange field is along the net magnetization in most cases, we discuss in Sec. IV-B a case where the exchange field opposes the net magnetization.

The oscillatory decay of $g_t^{\uparrow\downarrow}(d)$ modeled by Eq. (3) is illustrated in Fig. 5(c). Although Fig. 5(c) shows a case with $g_{t,0}^{\uparrow\downarrow}$ as a positive real quantity, $g_{t,0}^{\uparrow\downarrow}$ is generally complex. In particular, $\text{Re}[g_{t,0}^{\uparrow\downarrow}]$ and $\text{Im}[g_{t,0}^{\uparrow\downarrow}]$ represent the filtering and rotation [5,6], respectively, of spin current at the interface of the Cu spacer and the $Co_{100-x}Gd_x$ sink.

We incorporate Eq. (3) into the drift-diffusion model by Taniguchi *et al.* [10] that uses the boundary conditions applicable to our multilayer systems. Accordingly, the nonlocal Gilbert damping enhancement $\Delta\alpha$ due to spin decoherence in the spin sink is given by

$$\Delta\alpha = \frac{g\mu_B}{4\pi M_s t_F} \left( \frac{1}{\tilde{g}_1^{\uparrow\downarrow}} + \frac{1}{\tilde{g}_2^{\uparrow\downarrow}} \right)^{-1}, \quad (4)$$

where $\mu_B$ is the Bohr magneton; $g = 2.1$ is the gyromagnetic ratio, $M_s = 800$ kA/m is the saturation magnetization, and $t_F = 7$ nm is the thickness of the NiFe spin source. $\tilde{g}_1^{\uparrow\downarrow} = 16$ nm$^{-2}$ is the renormalized reflected spin-mixing conductance at the NiFe/Cu interface (see Appendix B). $\tilde{g}_2^{\uparrow\downarrow}$ is the renormalized reflected spin-mixing conductance at the Cu/$Co_{100-x}Gd_x$ interface, which depends on the $Co_{100-x}Gd_x$ spin sink thickness $d$ as

$$\tilde{g}_2^{\uparrow\downarrow} = \frac{\left(1 + \text{Re}[g_t^{\uparrow\downarrow}]\text{Re}[\eta] + \text{Im}[g_t^{\uparrow\downarrow}]\text{Im}[\eta]\right)}{\left(1 + \text{Re}[g_t^{\uparrow\downarrow}]\text{Re}[\eta] + \text{Im}[g_t^{\uparrow\downarrow}]\text{Im}[\eta]\right)^2 + \left(\text{Im}[g_t^{\uparrow\downarrow}]\text{Re}[\eta] - \text{Re}[g_t^{\uparrow\downarrow}]\text{Im}[\eta]\right)^2} \tilde{g}_r^{\uparrow\downarrow}, \quad (5)$$

where $\eta = (2e^2\rho l_+/h)\coth(d/l_+)$ with $l_+ = \sqrt{1/\lambda_{sf}^2 - i/\lambda_{dp}^2}$, $\rho$ is the resistivity of the spin sink, and $\tilde{g}_r^{\uparrow\downarrow}$ is the renormalized reflected mixing conductance at the Cu/ $Co_{100-x}Gd_x$ interface with $d \gg \lambda_{dp}$.

Given the rather large number of parameters in the model, some assumptions to constrain the fitting are required, as outlined in Appendix B. These assumptions leave us with three free fit parameters: $\lambda_{dp}$, $\mathrm{Re}[g_{t,0}^{\uparrow\downarrow}]$, and $\mathrm{Im}[g_{t,0}^{\uparrow\downarrow}]$. It is also possible to further reduce the number of free parameters by fixing $\mathrm{Re}[g_{t,0}^{\uparrow\downarrow}]$, particularly near the compensation composition as explained in Appendix C. Qualitatively similar results are obtained with $\mathrm{Re}[g_{t,0}^{\uparrow\downarrow}]$ free or fixed.

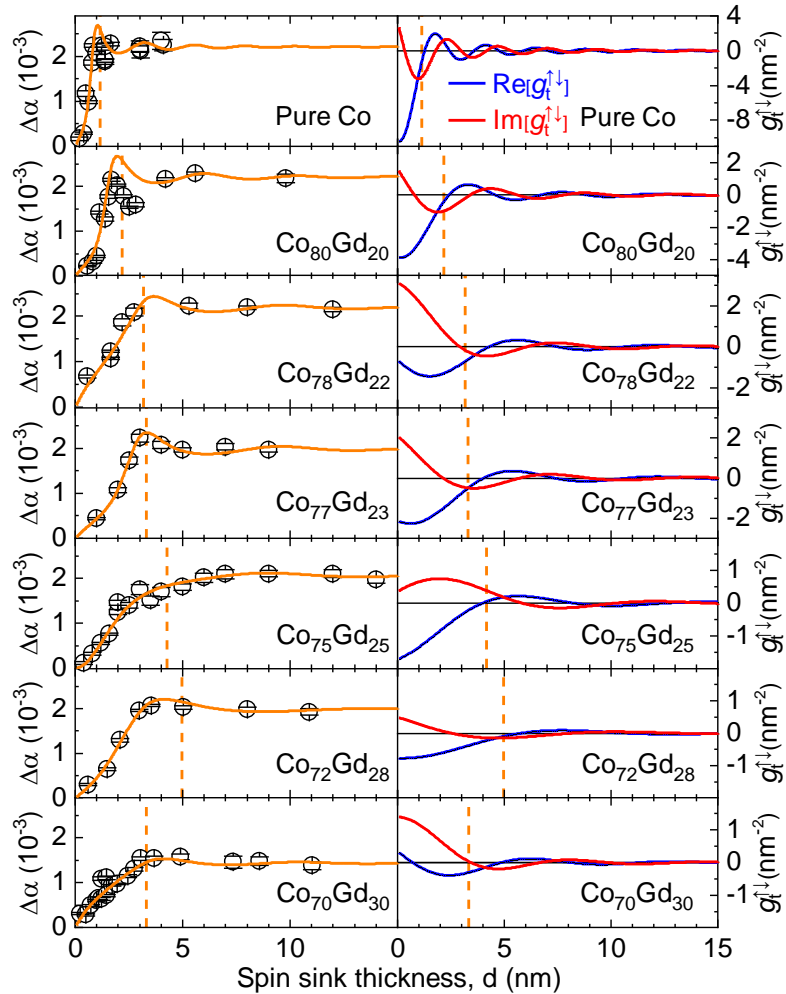

**FIG. 6.** Left column: spin sink thickness dependence of nonlocal damping enhancement Δα. The solid curve indicates the fit using the modified drift-diffusion model. Right column: the real and imaginary parts of the transmitted spin-mixing conductance $g_t^{\uparrow\downarrow}$ derived from the fit. The vertical dashed line indicates the spin dephasing length $\lambda_{dp}$.

The fit results using the modified drift-diffusion model are shown as solid curves in the left column of Fig. 6. These curves adequately reproduce the $d$ dependence of Δ$α$ for all spin sink compositions. We also show in the right column of Fig. 6 the $d$ dependence of $g_t^{\uparrow\downarrow}$, illustrating the net precession and decay of the transverse spin current.

With the modeled results in Fig. 6, the overshoot in Δ$α$ versus $d$ can now be attributed to the precession of the transverse spin current [9,44,77]. For a certain magnetic spin sink thickness, the polarization of the spin current leaving the sink is opposite to that of the spin current entering the spin sink. Since the difference between the leaving and entering spin currents is related to the spin angular momentum transferred to the magnetic order, the spin transfer – manifesting as Δ$α$ here – can be enhanced [9,44,77] compared to when the spin current is completely dephased for $d \gg \lambda_{dp}$.

We acknowledge that our assumptions in the modified drift-diffusion model do not necessarily capture all phenomena that could impact spin transport in a quantitative manner. For example, there remain unresolved questions regarding the relationship between resistivity $ρ$ and spin-flip length $\lambda_{sf}$, as well as the role of the thickness dependent compensation composition on dephasing, which warrant further studies in the future. Nevertheless, as discussed in Sec. IV, our approach reveals salient features of spin dephasing in compensated FIMs, as well as the upper bound of the transverse spin coherence length in FIM CoGd.

## IV. DISCUSSION
### A. Compositional Dependence of Spin Transport

Figure 7 summarizes the main finding of our work, namely the relationship between the magnetic compensation (Fig. 7(a)) and the spin transport parameters (Fig. 7(b-d)) of FIM $Co_{100-x}Gd_x$. We also refer the readers to Fig. 16 in Appendix C, which shows that qualitatively similar results are obtained when $\mathrm{Re}[g_{t,0}^{\uparrow\downarrow}]$ is treated as a fixed parameter.

We first discuss the composition dependence of $\lambda_{dp}$, summarized in Fig. 7(b), which is derived from the modified drift-diffusion model (vertical dashed lines in Fig. 6). The results in Fig. 7(b) show a peak value of $\lambda_{dp} ≈ 5$ nm at $x ≈ $ 25-28, which is close to the magnetic compensation composition window ($x ≈ $ 22-25, the shaded region in Fig. 7). It is worth noting that $\lambda_{dp}$ is maximized on the somewhat Gd-rich side of the compensation composition window. We attribute this observation to the stronger contribution to spin dephasing from the Co sublattice, as discussed more in detail in Sec. IV-B.

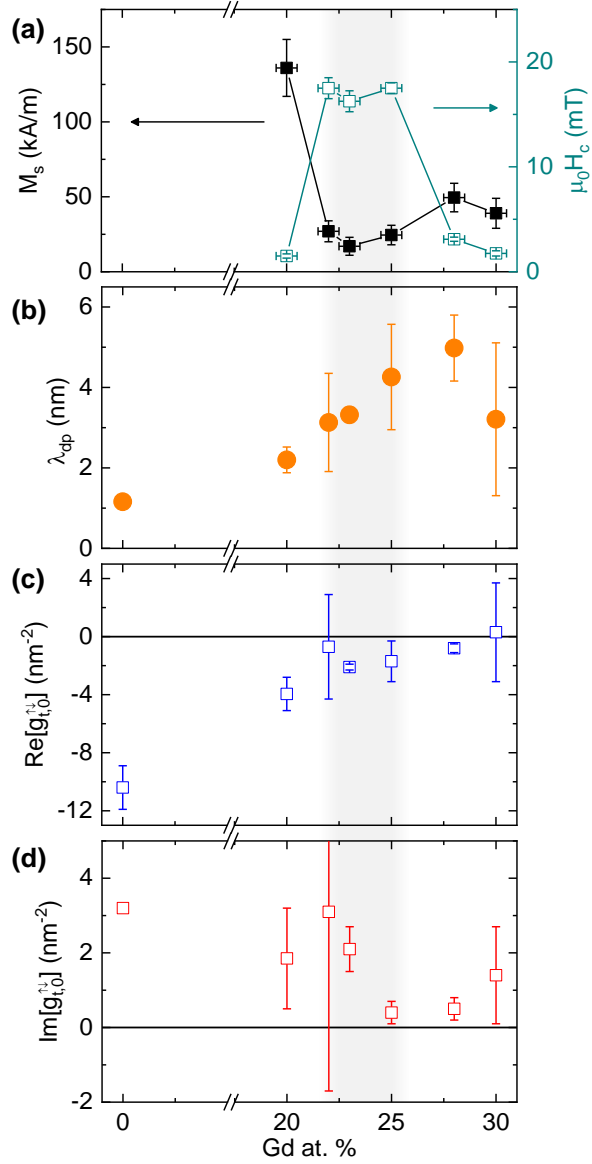

**FIG. 7.** (a) Static magnetic properties (saturation magnetization $M_s$ and coercive field $H_c$) reproduced from Fig. 2, (b) spin dephasing length $\lambda_{dp}$, and the (c) real and (d) imaginary parts of the interfacial contribution of the transmitted spin-mixing conductance $g_{t,0}^{\uparrow\downarrow}$ versus Gd content in the spin sink. The shaded region indicates the window of composition corresponding to magnetic compensation. The error bars in (b-d) show the 95% confidence interval.

Our results (Fig. 7(b)) indicate up to a factor of ≈5 enhancement in $\lambda_{dp}$ for nearly compensated FIMs compared to FM Co. This observation is qualitatively consistent with the theoretical prediction that antiferromagnetic order mitigates the decoherence of transverse spin current [14–18]. In a nearly compensated FIM CoGd spin sink, the alternating Co and Gd moments of approximately equal magnitude partially cancel the dephasing of the propagating spins. This scenario, illustrated in Fig. 1(b), is corroborated by our tight-binding calculations assuming coherent ballistic transport (see Appendix D). Transverse spin current in compensated CoGd is therefore able to remain coherent over a longer distance than in FMs. We note, however, that the spin current decoheres within a finite length scale in any real materials, due to the imperfect suppression of dephasing and the presence of incoherent scattering.

We now comment on the compositional dependence of $\mathrm{Re}[g_{t,0}^{\uparrow\downarrow}]$ and $\mathrm{Im}[g_{t,0}^{\uparrow\downarrow}]$, particularly in the vicinity of magnetic compensation. Our calculations in Appendix C predict $\mathrm{Re}[g_{t,0}^{\uparrow\downarrow}]$ to be only weakly dependent on the net exchange splitting $k_\Delta$ of the magnetic spin sink. Our experimental results (Fig. 7(c)) are in qualitative agreement with this prediction, as $\mathrm{Re}[g_{t,0}^{\uparrow\downarrow}]$ takes a roughly constant value near magnetic compensation.

By contrast, the same calculations in Appendix C predict a quadratic dependence of $\mathrm{Im}[g_{t,0}^{\uparrow\downarrow}]$ on $k_\Delta$. Specifically, $\mathrm{Im}[g_{t,0}^{\uparrow\downarrow}]$ converges to zero as the net exchange $k_\Delta$ approaches zero (i.e., magnetic sublattices approaching compensation). Our experimental results indeed show a minimum in $\mathrm{Im}[g_{t,0}^{\uparrow\downarrow}]$ when $\lambda_{dp}$ is maximized. We remark that $\mathrm{Im}[g_{t,0}^{\uparrow\downarrow}]$ is related to how much the polarization of the spin current rotates upon entering the magnetic spin sink [5,6]. When the magnetic sublattices are nearly compensated, the spin current sees a nearly canceled net exchange field such that the spin rotation (precession) is suppressed. Therefore, both the reduction of $\mathrm{Im}[g_{t,0}^{\uparrow\downarrow}]$ and the enhancement of $\lambda_{dp}$ arise naturally from the cancellation of the net exchange field. Our results in Figs. 6 and 7 consistently point to the suppression of spin-current dephasing enabled by antiferromagnetic order.

It is important to note that FIM TM-RE alloys in general are amorphous with no long-range structural order. Instead of the simple layer-by-layer alternating order illustrated in Fig. 1(b), the TM and RE atoms are expected to be arranged in a rather disordered fashion. Considering that disorder and electronic scattering tend to quench transverse spin coherence [18,23,24], it is remarkable that such amorphous FIMs permit extended $\lambda_{dp}$ at all. We speculate the observed enhancement of transverse spin coherence is enabled by short-range ordering of Co and Gd atoms, e.g., finite TM-TM and RE-RE pair correlations in the film plane (and TM-RE pair correlation out of the film plane) as suggested by prior reports [18,80].

**B. Distinct Influence of the Sublattices on Spin Dephasing**

Our results (Fig. 7) indicate that magnetic compensation is achieved in the $Co_{100-x}Gd_x$ composition range of $x \approx$ 22-25, while $\lambda_{dp}$ is maximized (and $\text{Im}[g_{t,0}^{\uparrow\downarrow}]$ is minimized) at $x \approx$ 25-28. This observation deviates from the simple expectation (Fig. 1(b)) that spin dephasing is suppressed when the two sublattices are compensated. Here, we discuss why $\lambda_{dp}$ should be maximized at a more Gd-rich spin sink composition than the magnetic compensation composition.

First, it should be recalled that the magnetic compensation composition is dependent on the FIM thickness. Our magnetometry data (Fig. 2(b-g)) indicate that the magnetic compensation composition becomes more Gd-rich with decreasing CoGd thickness $d$. However, the CoGd thickness $d = 5$ nm shown in Fig. 2(f,g), from which the compensation composition is deduced, is close to the estimated maximum $\lambda_{dp}$ of ≈ 5 nm. Therefore, the thickness dependence of the compensation composition alone does not explain the maximum $\lambda_{dp}$ on the Gd-rich side of magnetic compensation. We consider an alternative explanation below.

Generally, it might be expected that transverse spin current interacts more strongly with the TM Co magnetization (from the spin-split itinerant 3*d* bands near the Fermi level) [50,81] than the RE Gd magnetization (primarily from the localized 4*f* levels ≈7-8 eV below the Fermi level [82,83]). This is analogous to magnetotransport effects dominated by itinerant 3*d* band magnetism in TM-RE FIMs [60,84]. If the interaction of transverse spin current with the Co sublattice is stronger [81], more Gd would be needed to compensate spin dephasing. Thus, the greater contribution of the Co sublattice to dephasing could explain why $\lambda_{dp}$ is maximized at a more Gd-rich composition than the magnetic compensation composition.

We observe additional evidence for the stronger interaction with the TM Co sublattice by inspecting the dependence of $g_t^{\uparrow\downarrow}$ on spin sink thickness $d$. The relevant results can be seen in the right column of Fig. 6, but for the sake of clarity, we highlight $g_t^{\uparrow\downarrow}$ vs $d$ for a selected few CoGd compositions near magnetic compensation in Fig. 8. For most spin sink compositions (e.g., Fig. 8(a),(c)), the results are consistent with the straightforward scenario: the net exchange field, felt by the transverse spin current, points along the net magnetization. However, the $Co_{75}Gd_{25}$ spin sink (Fig. 8(b)), which is on the Gd-rich side of the magnetic compensation composition, exhibits a qualitatively different phase shift in $\text{Im}[g_t^{\uparrow\downarrow}]$. This noteworthy result is consistent with the transverse spin precessing about a net exchange field that *opposes* the net magnetization[3]. In other words, the net magnetization in $Co_{75}Gd_{25}$ is dominated by the Gd magnetization (from 4*f* orbitals far below the Fermi level), but the net exchange field points along the Co magnetization (from 3*d* bands near the Fermi level). The retrograde spin precession in $Co_{75}Gd_{25}$ suggests that transverse electronic spin current interacts preferentially with the itinerant 3*d* TM magnetism.

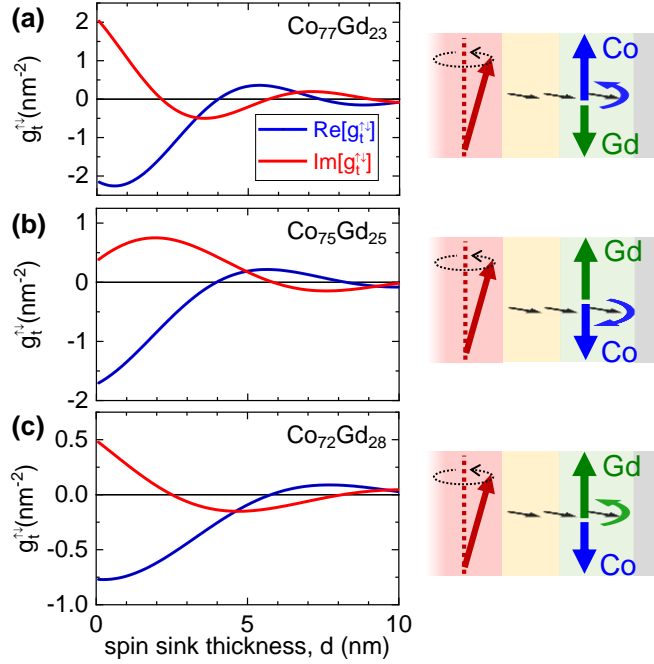

**FIG. 8.** Transmitted spin-mixing conductance $g_t^{\uparrow\downarrow}$ versus spin sink thickness $d$ near the magnetic compensation composition, derived from the modified drift-diffusion model fit (Fig. 6). Note the phase shift in $\text{Im}[g_t^{\uparrow\downarrow}]$ for $Co_{75}Gd_{25}$ (b). The cartoons on the right illustrate the net precession direction of the transverse spin polarization in each CoGd spin sink. $Co_{75}Gd_{25}$ exhibits retrograde precession, opposite to the precession direction in the other spin sink compositions.

---

[3] The modeled curves for $\text{Re}[g_t^{\uparrow\downarrow}]$ and $\text{Im}[g_t^{\uparrow\downarrow}]$ in Figs. 6 and 8 for $Co_{75}Gd_{25}$ showing retrograde precession are obtained by taking the negative sign between the real and imaginary terms in Eq. (3). Adequate fits could also be obtained by fixing the sign between the real and imaginary terms to be positive, but this would require the signs of $\text{Re}[g_{t,0}^{\uparrow\downarrow}]$ and $\text{Im}[g_{t,0}^{\uparrow\downarrow}]$ for $Co_{75}Gd_{25}$ to be opposite to those of the other CoGd compositions. Since a sign flip in $\text{Re}[g_{t,0}^{\uparrow\downarrow}]$ or $\text{Im}[g_{t,0}^{\uparrow\downarrow}]$ with respect to CoGd composition is not expected (according to our calculations in Appendix C), we conclude that retrograde precession is the physically reasonable scenario for the $Co_{75}Gd_{25}$ sink.

We also comment on the possible role of angular momentum compensation in FIM CoGd. As noted in Sec. II, angular momentum compensation composition is slightly more Co-rich than the magnetic compensation composition. While angular momentum compensation is key to fast antiferromagnetic-like dynamics in FIMs [53,55], it is evidently unrelated to the maximum $\lambda_{dp}$ on the slightly Gd-rich side of the magnetic compensation composition. Rather, we conclude that $\lambda_{dp}$ in FIMs is governed by the effective net exchange field, where the TM sublattice can play a greater role than the RE sublattice.

**C. Comparison with a Prior Report of Long Transverse Spin Coherence in Ferrimagnetic Metals**

Our results (e.g., Figs. 6-8) point to mitigation of spin decoherence in nearly compensated FIM CoGd. However, we do not observe evidence for a transverse spin coherence length in excess of 10 nm, which was recently reported for CoTb [18]. Owing to the weaker spin-orbit coupling in CoGd than in CoTb, one might expect longer length scales for spin dephasing (more collinear antiferromagnetic order) and spin diffusion (less spin-flip scattering) in CoGd than in CoTb. We discuss possible reasons as to why the maximum coherence length in CoGd in our present study is significantly shorter than that reported in CoTb by Ref. [18].

A plausible factor is the difference in experimental method for deducing the coherence length. Yu *et al.* in Ref. [18] utilize spin-galvanic measurements on Co/Cu/CoTb/Pt stacks: FMR in the Co layer pumps a spin current presumably through the CoTb spacer and generates a lateral dc voltage from the inverse spin-Hall effect in the Pt detector [85]. A finite dc voltage is detected for a range of CoTb alloy spacer thicknesses up to 12 nm, interpreted as evidence that the spin current propagates from Co to Pt even with >10 nm of CoTb in between. However, there could be coexisting voltage contributions besides the spin-to-charge conversion in the Pt detector. For example, spin scattering in the CoTb layer could yield an additional inverse spin-Hall effect [86], i.e., the reciprocal of the strong spin-orbit torque reported in CoTb [87]. Furthermore, the FMR-driven spin-galvanic measurement could pick up spin rectification [31–33] and thermoelectric voltages [34,35] from the dynamics of the FM Co layer, which might be challenging to disentangle from the inverse spin-Hall effect in Pt. While a number of control experiments are performed in Ref. [18] to rule out artifacts, it is possible that some spurious effects are not completely suppressed. Figures 4(d) and 5(c) in Ref. [18] show that the spin-galvanic signal drops abruptly with the inclusion of a finite thickness of CoTb spacer and remains constant up to CoTb thickness >10 nm. This trend, at odds with the expected gradual attenuation of spin current with CoTb thickness, may arise from other mechanisms unrelated to spin transmission through CoTb.

In our present study, the spin pumping method measures the nonlocal damping $\Delta\alpha$ that is attributed to spin-current decoherence in the spin sink (see Sec. III-A). This method does not involve any complications from spin-galvanic signals. It still might be argued, however, that our method does not necessarily allow for precise, straightforward quantification of spin transport due to the large number of parameters involved in modeling (see Appendix B). Nevertheless, even if there are quantitative errors in our modeling, it is incontrovertible that $\Delta\alpha$ saturates (i.e., the transverse spin current pumped into CoGd decoheres) within a length scale well below 10 nm. Thus, our results indicate that the spin coherence length in CoGd does *not* exceed 10 nm.

There may be other factors leading to the discrepancy between our study and Ref. [18]. The CoTb films in Ref. [18] may have a higher degree of layer-by-layer ordering than our CoGd films. This appears unlikely, since CoTb "multilayers" (grown by alternately depositing Co and Tb) and CoTb "alloys" (grown by simultaneously depositing Co and Tb) exhibit essentially the same CoTb thickness dependence of spin-galvanic signal [18]. Another possibility is that magnons, rather than spin-polarized conduction electrons, are responsible for the remarkably long spin coherence through CoTb. If this were the case, it is yet unclear how CoTb might exhibit a longer magnon coherence length than CoGd, or how the spin-galvanic method in Ref. [18] might be more sensitive to magnon spin transport than the nonlocal damping method in our present study. A future study that directly compares CoTb and CoGd spin sinks (e.g., with the nonlocal damping method) may verify whether transverse spin currents survive over >10 nm in ferrimagnetic alloys with strong spin-orbit coupling.

**D. Possible Implications for Spintronic Device Applications**

The dephasing length $\lambda_{dp}$ of transverse electronic spin current fundamentally impacts spin torque effects in a magnetic metal [5,6]. Specifically, the transverse spin angular momentum lost by the spin current is transferred to the magnetization, thereby giving rise to a spin torque within a depth of order $\lambda_{dp}$ from the surface of the magnetic metal. Due to the short $\lambda_{dp} \approx 1$ nm, the spin torque is more efficient in a thinner FM layer, which comes at the expense of reduced thermal stability of the stored magnetic information. The longer $\lambda_{dp}$ in compensated FIMs (or AFMs) may enable efficient spin torques in thicker, more thermally stable magnetic layers for high-density nonvolatile memory devices [18,52]. Another practical benefit of extended $\lambda_{dp}$ is that it facilitates tuning the magnetic layer thickness to enhance spin torques [9,44,77]. We further emphasize that the increase of $\lambda_{dp}$ is evident even in amorphous FIMs. This suggests that optimally engineered FIMs or AFMs (e.g., with high crystallinity) could exhibit much longer $\lambda_{dp}$, potentially yielding spin torque effects that are qualitatively distinct from those in FMs [14].

In addition to estimating the dephasing length scale in FIMs, our study highlights the roles of the chemically distinct sublattices on spin dephasing. From conventional spin torque measurements, it is generally difficult to deduce whether a spin current in a TM-RE FIM interacts more strongly with a particular sublattice [54]. Our results (as discussed in Sec. IV-B) imply that the TM sublattice contributes more strongly to the dephasing of transverse spin current – and hence to spin torques. The greater role of the TM sublattice over the RE sublattice could be crucial for engineering spin torque effects in compensated TM-RE FIMs – e.g., for enabling ultrafast (sub-THz-range) spin torque oscillators [88].

**V. CONCLUSION**

In summary, we have utilized broadband FMR spin pumping to estimate the dephasing length $\lambda_{dp}$ of transverse spin current in ferrimagnetic CoGd alloys across the compensation point. We obtain a maximum of $\lambda_{dp} \approx 5$ nm in nearly compensated CoGd, consistent with the antiferromagnetic order mitigating the decoherence (dephasing) of transverse spin current. The observed maximum $\lambda_{dp}$ constitutes a factor of ≈4-5 enhancement compared to that for ferromagnetic metals. On the other hand, we do not find evidence for $\lambda_{dp}$ in excess of 10 nm in ferrimagnetic alloys reported in a recent study [18]. Despite this quantitative difference, our results suggest that partial spin rephasing by antiferromagnetic order – i.e., analogous to the spin-echo scheme to counter spin decoherence – is indeed operative even in disordered ferrimagnetic alloys at room temperature. Moreover, our results suggest that transverse spin current interacts more strongly with the itinerant Co sublattice than the localized Gd sublattice in nearly compensated CoGd. The spin rephasing effect and the sublattice dependent interaction could impact spin torques in ferrimagnetic alloys, with possible applications in fast spintronic devices. Our finding also points to the possibility of further extending transverse spin coherence in structurally pristine antiferromagnetic metals, thus opening a new avenue for fundamental studies of spin transport in magnetic media.

Acknowledgements

This research was funded in part by 4-VA, a collaborative partnership for advancing the Commonwealth of Virginia, as well as by the ICTAS Junior Faculty Award and the NSF Grant No. DMR-2003914. We thank Jean J. Heremans for helpful discussions.

## APPENDIX A: FMR MEASUREMENT METHOD

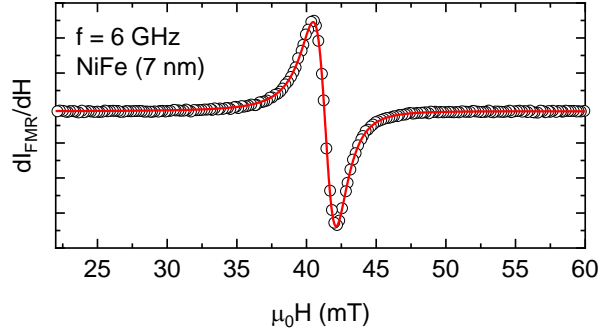

**FIG. 9.** FMR spectrum of $Ni_{80}Fe_{20}(7)/Cu(4)$ at 6 GHz. The red curve in the bottom panel represents the fit using Eq. (A1).

FMR spectra are acquired using a broadband spectrometer, with each sample placed on a coplanar waveguide (film side down) and magnetized in-plane (with a quasi-static magnetic field, maximum value ~1 T, from an electromagnet). The quasi-static field is swept while fixing the frequency of the microwave field (transverse to the quasi-static field) to acquire the resonance spectrum. A radio-frequency diode and lock-in amplifier (with 700 Hz modulation field as the reference) is used to detect the signal, which is recorded as the derivative of the microwave power absorption with respect to the applied field, as shown in Fig. 9. To obtain the half-width-at-half-maximum FMR linewidth $\Delta H$, the measured signal is fit with the derivative of the sum of symmetric and antisymmetric Lorentzian functions [89],

$$\frac{dI_{FMR}}{dH} = A \frac{2(H-H_{res})\Delta H}{[(H-H_{res})^2 + (\Delta H)^2]^2} + S \frac{(H-H_{res})^2 - (\Delta H)^2}{[(H-H_{res})^2 + (\Delta H)^2]^2}, \qquad (A1)$$

where $H_{res}$ is the resonance field and the coefficients $A$ and $S$ are the proportionality factors for the antisymmetric and symmetric terms, respectively.

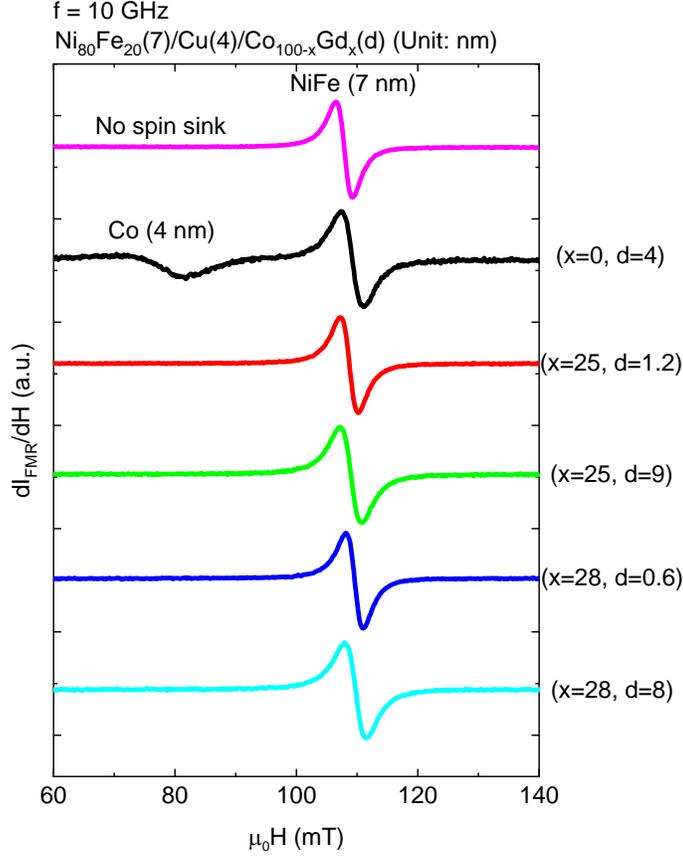

**FIG. 10.** FMR spectra for $Ni_{80}Fe_{20}$ (7 nm)/Cu (4 nm)/$Co_{100-x}Gd_x$ ($d$ nm) with no spin sink (magenta) and with spin sink layer of pure Co (black), Gd 25% (red and green) and Gd 28% (blue and light blue). NiFe and Co exhibit well-separated FMR. The different thicknesses for Gd 25 and 28% was selected to show two different transport regimes, i.e., below/above the $\lambda_{dp}$ found from Fig. 6 in the main text. No FMR signal attributable to $Co_{100-x}Gd_x$ in our samples (20 ≤ x ≤ 30) is detected.

## APPENDIX B: ASSUMPTIONS IN THE MODIFIED DRIFT-DIFFUSION MODEL

Given the rather large number of parameters in the model ($\tilde{g}_1^{\uparrow\downarrow}$, $\tilde{g}_r^{\uparrow\downarrow}$, $\rho$, $\lambda_{sf}$, $\lambda_{dp}$, $\text{Re}[g_{t,0}^{\uparrow\downarrow}]$, and $\text{Im}[g_{t,0}^{\uparrow\downarrow}]$), some assumptions are required to constrain the modified drift-diffusion model (Sec. III-B). Our assumptions are as follow:

(1) The renormalized reflected spin-mixing conductance (accounting for the Sharvin conductance of Cu [79]) is set equal for the $Ni_{80}Fe_{20}$/Cu and Cu/Co interfaces, i.e., $\tilde{g}_1^{\uparrow\downarrow} = \tilde{g}_r^{\uparrow\downarrow}$. Here, $\tilde{g}_r^{\uparrow\downarrow}$ is the renormalized reflected spin-mixing conductance for the Cu/Co interface, $\tilde{g}_2^{\uparrow\downarrow}$, in the limit of $d \gg \lambda_{dp}$. We compute $\tilde{g}_1^{\uparrow\downarrow}$ from a modified form of Eq. (4),

$$\Delta\alpha_{sat}^{Co} = \frac{g\mu_B}{4\pi M_s t_F}\left(\frac{2}{\tilde{g}_1^{\uparrow\downarrow}}\right)^{-1}, \quad (B1)$$

where $\Delta\alpha_{sat}^{Co} = 0.0022$ is the average of $\Delta\alpha$ for Co spin sink thicknesses $d \geq 3$ nm (large enough that the transverse spin current is essentially completely dephased). The renormalized spin-mixing conductance $\tilde{g}_1^{\uparrow\downarrow}$ at the Ni$_{80}$Fe$_{20}$/Cu interface is found to be 16 nm$^{-2}$.

(2) For each ferrimagnetic Co$_{100-x}$Gd$_x$ spin sink composition, we compute $\tilde{g}_r^{\uparrow\downarrow}$, shown in Eq. (5)) via

$$\Delta\alpha_{sat}^{CoGd} = \frac{g\mu_B}{4\pi M_s t_F}\left(\frac{1}{\tilde{g}_1^{\uparrow\downarrow}} + \frac{1}{\tilde{g}_r^{\uparrow\downarrow}}\right)^{-1}, \quad (B2)$$

where $\Delta\alpha_{sat}^{CoGd}$ is the average of $\Delta\alpha$ for the samples with the three largest spin sink thicknesses, where $\Delta\alpha$ is essentially saturated. The values of $\tilde{g}_r^{\uparrow\downarrow}$ are summarized in Fig. 11.

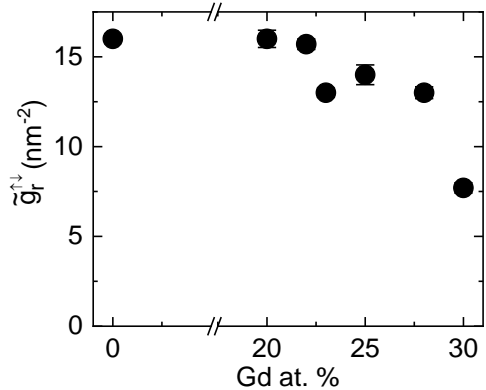

**FIG. 11.** Composition dependence of the renormalized reflected mixing conductance $\tilde{g}_r^{\uparrow\downarrow}$ at the Cu/ Co$_{100-x}$Gd$_x$ interface with the Co$_{100-x}$Gd$_x$ spin sink thickness $d \gg \lambda_{dp}$.

(3) We set $\lambda_{sf} = 10$ nm, which is of the same order as the reported spin diffusion length in Co [4]. We further note that if the claim of $\lambda_c > 10$ nm for ferrimagnets in Ref. [18] were correct, $\lambda_{sf}$ would necessarily need to be $> 10$ nm. This relatively long $\lambda_{sf}$ is equivalent to assuming quasi-ballistic spin transport in the spin sink, such that spin dephasing (rather than scattering) governs the transverse spin coherence length, $\lambda_{dp} < \lambda_{sf}$. Our quantitative results of $\lambda_{dp}$ are essentially unaffected if $\lambda_{sf} \gtrsim 4$ nm (e.g., Fig. 12), while some variation can be seen in $\text{Im}[g_{t,0}^{\uparrow\downarrow}]$ with $\lambda_{sf}$.

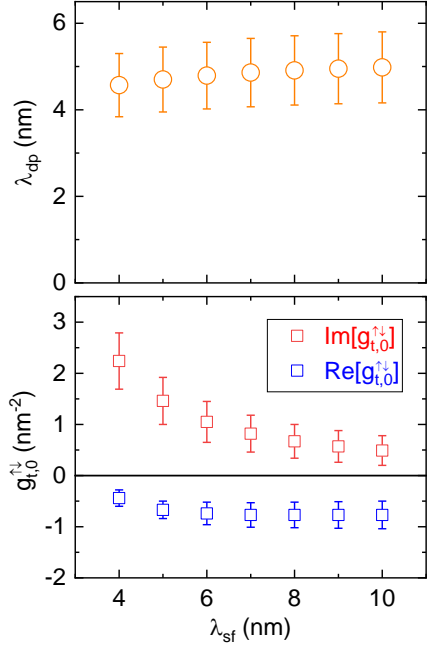

**FIG. 12.** Dependence of the spin dephasing length $\lambda_{dp}$ and interfacial transmitted spin-mixing conductance $g_{t,0}^{\uparrow\downarrow}$ of $Co_{72}Gd_{28}$ on the spin-flip length $\lambda_{sf}$ in the modified drift-diffusion model.

(4) For the sake of simplicity, we assume constant values of $\lambda_{sf}$ and $\rho$ for a given spin-sink composition (although in general, $\lambda_{sf}$ may depend on the resistivity $\rho$ of the spin sink, which in turn can depend on the spin-sink thickness). In particular, we set $\rho$ at the values obtained from four-point resistivity measurements of 10-nm-thick $Co_{100-x}Gd_x$ films (Fig. 13); the results are summarized in the figure below. (The exception was pure Co where $\rho$ was set at $1.0 \times 10^{-6}$ $\Omega$m, since using the measured resistivity $\rho = 0.3 \times 10^{-6}$ $\Omega$m yielded an unphysically large $|g_{t,0}^{\uparrow\downarrow}|$ of $\gg 10$ nm$^{-2}$. The higher effective resistivity for Co is possibly justified considering that $\Delta\alpha$ saturates at $d \approx 1$ nm, where the resistivity is likely much higher than $0.3 \times 10^{-6}$ $\Omega$m due to surface scattering.) We remark that the uncertainty in the spin-sink thickness dependence of $\rho$ and $\lambda_{sf}$ may result in a systematic error in quantifying $\lambda_{dp}$; further detailed studies are warranted to elucidate the relationship between electronic and spin transport in ferrimagnets. Nevertheless, our approach is sufficient for semi-quantitative examination of $\lambda_{dp}$ as a function of CoGd composition. It is incontrovertible that FIM CoGd has a significantly longer transverse coherence length than FM Co and that this coherence length (albeit well below 10 nm) shows a maximum near the magnetic compensation.

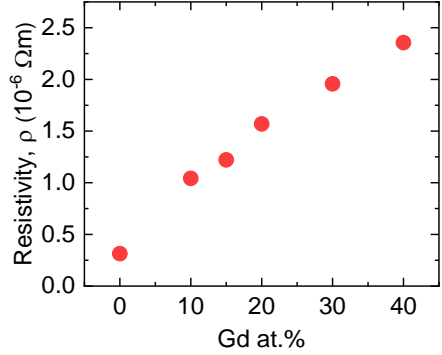

**FIG. 13.** Resistivity $\rho$ with varying Gd at.% for SiO$_x$(thermally oxidized)/Co$_{100-x}$Gd$_x$(10)/TiO$_x$(3) (unit: nm). The resistivities of intermediate compositions between $x$ = 20 and 30 are derived via linear interpolation.

(5) We assume $\lambda_{dp}$ to be a constant parameter for each spin-sink composition, i.e., independent of the spin sink thickness. We note that there could be a deviation from this assumption, considering that the magnetic compensation composition (hence net exchange splitting) evidently depends on the thickness of the ferrimagnetic spin sink, as seen in our magnetometry results (Fig. S3).

We also add a few remarks regarding the details of our fitting protocol. In fitting the experimental data $\Delta\alpha$ versus $d$ using the modified drift-diffusion model (Fig. 6, left column), we assign a weight to each data point that is inversely proportional to the square of the error bar for $\Delta\alpha$ (from the linear fit of linewidth versus frequency). We fix $\mathrm{Im}[g_{t,0}^{\uparrow\downarrow}] = 0.2\tilde{g}_r^{\uparrow\downarrow}$ for pure Co, similar to what is suggested by Zwierzycki *et al.* [78]; for Co$_{80}$Gd$_{20}$, we impose the constraint $0.2\tilde{g}_r^{\uparrow\downarrow} < |g_{t,0}^{\uparrow\downarrow}| < 0.3\tilde{g}_r^{\uparrow\downarrow}$, since having the magnitudes of $\mathrm{Re}[g_{t,0}^{\uparrow\downarrow}]$ and $\mathrm{Im}[g_{t,0}^{\uparrow\downarrow}]$ as free parameters resulted in unphysically large $|g_{t,0}^{\uparrow\downarrow}|$ with large error bars.

## APPENDIX C. CONSTANT $\mathrm{Re}[g_{t,0}^{\uparrow\downarrow}]$ NEAR MAGNETIC COMPENSATION

While the assumptions outlined in Appendix B results in three free fit parameters in the drift-diffusion model ($\lambda_{dp}$, $\mathrm{Re}[g_{t,0}^{\uparrow\downarrow}]$, and $\mathrm{Im}[g_{t,0}^{\uparrow\downarrow}]$), it is possible to reduce the number of free parameters further by fixing $\mathrm{Re}[g_{t,0}^{\uparrow\downarrow}]$. Here, we provide a physical justification for setting $\mathrm{Re}[g_{t,0}^{\uparrow\downarrow}]$ constant, particularly near the magnetic compensation composition.

We consider the simplest non-magnet and ferromagnet interface. In the nonmagnetic metal, $k_x = \sqrt{k_F^2 - \kappa^2}$, and in the ferromagnet, $k_x^\sigma = \sqrt{k_F^2 + \sigma k_\Delta^2 - \kappa^2}$. Here $\kappa$ is the momentum lying in the plane of the interface and $k_\Delta = \frac{\sqrt{2m\Delta}}{\hbar}$ quantifies the s-d exchange. The transmission coefficient $t_\sigma$ for spin $\sigma$ reads

$$t_\sigma = \frac{2k_x}{k_x^\sigma + k_x}, \quad (C1)$$

and consequently, the transmitted mixing conductance at the interface is

$$g_{t,0}^{\uparrow\downarrow} = \int \frac{d^2\kappa}{4\pi^2} t_\uparrow (t_\downarrow)^* . \quad (C2)$$

Figure 14 displays the dependence of $t_\uparrow (t_\downarrow)^*$ as a function of the in-plane momentum $\kappa^2$. The real part is given in blue and the imaginary part is given in red. The first interesting feature is that all incident states contribute to $\mathrm{Re}[g_{t,0}^{\uparrow\downarrow}]$ whereas only states with incidence larger than $\kappa^2 \geq k_F^2 - k_\Delta^2$ (rather close to grazing incidence) contribute to $\mathrm{Im}[g_{t,0}^{\uparrow\downarrow}]$.

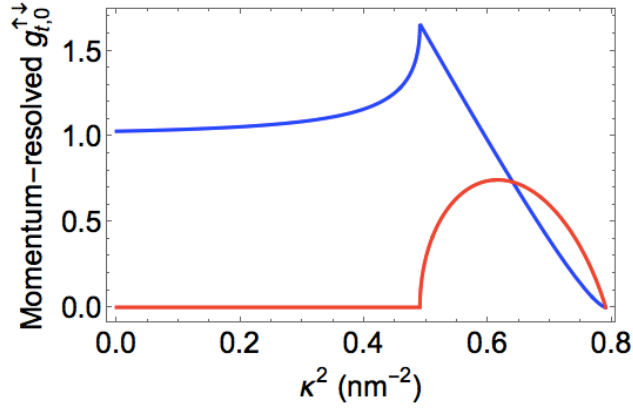

**FIG. 14.** Momentum-resolved $g_{t,0}^{\uparrow\downarrow}$ versus the in-plane momentum $\kappa^2$. $\mathrm{Re}[g_{t,0}^{\uparrow\downarrow}]$ in blue and $\mathrm{Im}[g_{t,0}^{\uparrow\downarrow}]$ in red.

Upon integration over the Fermi surface, the real and imaginary parts of the interfacial transmitted mixing conductance reduce to $\mathrm{Re}[g_{t,0}^{\uparrow\downarrow}] \approx \frac{k_F^2}{2\pi}$ and $\mathrm{Im}[g_{t,0}^{\uparrow\downarrow}] = \frac{k_\Delta^2}{4\pi}$. In other words, the real part is mostly independent of the exchange splitting, whereas the imaginary part is directly proportional to it (in fact, it is quadratic). This behavior is reported in Fig. 15. When varying the content of Gd in the vicinity of the magnetic compensation composition, it is therefore reasonable to assume that $\mathrm{Re}[g_{t,0}^{\uparrow\downarrow}]$ is constant whereas $\mathrm{Im}[g_{t,0}^{\uparrow\downarrow}]$ varies.

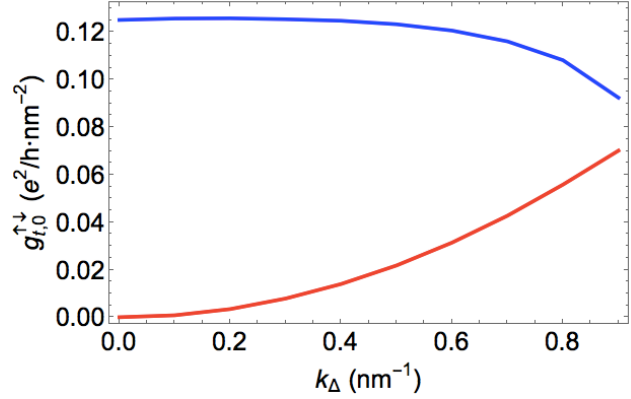

**FIG. 15.** Interfacial transmitted mixing conductance versus the exchange splitting $k_\Delta$. $\text{Re}[g_{t,0}^{\uparrow\downarrow}]$ in blue and $\text{Im}[g_{t,0}^{\uparrow\downarrow}]$ in red.

Figure 16 compares the fitting results of our experimental data with free $\text{Re}[g_{t,0}^{\uparrow\downarrow}]$ (i.e., same as Fig. 7) and with fixed $\text{Re}[g_{t,0}^{\uparrow\downarrow}]$. In Fig. 16(b), $\text{Re}[g_{t,0}^{\uparrow\downarrow}]$ is fixed at a constant magnitude of 1 nm$^{-2}$ for all samples, except for Co and Co$_{80}$Gd$_{20}$ where larger $\text{Re}[g_{t,0}^{\uparrow\downarrow}]$ (e.g., 8 and 4 nm$^{-2}$, respectively) is needed to obtain adequate fit curves. We observe that the qualitative conclusion is unaffected by whether or not $\text{Re}[g_{t,0}^{\uparrow\downarrow}]$ is treated as a free parameter: $\lambda_{dp}$ is maximized (and $\text{Im}[g_{t,0}^{\uparrow\downarrow}]$ is minimized) close to – or, more specifically, on the slightly Gd-rich side of – the magnetic compensation composition.

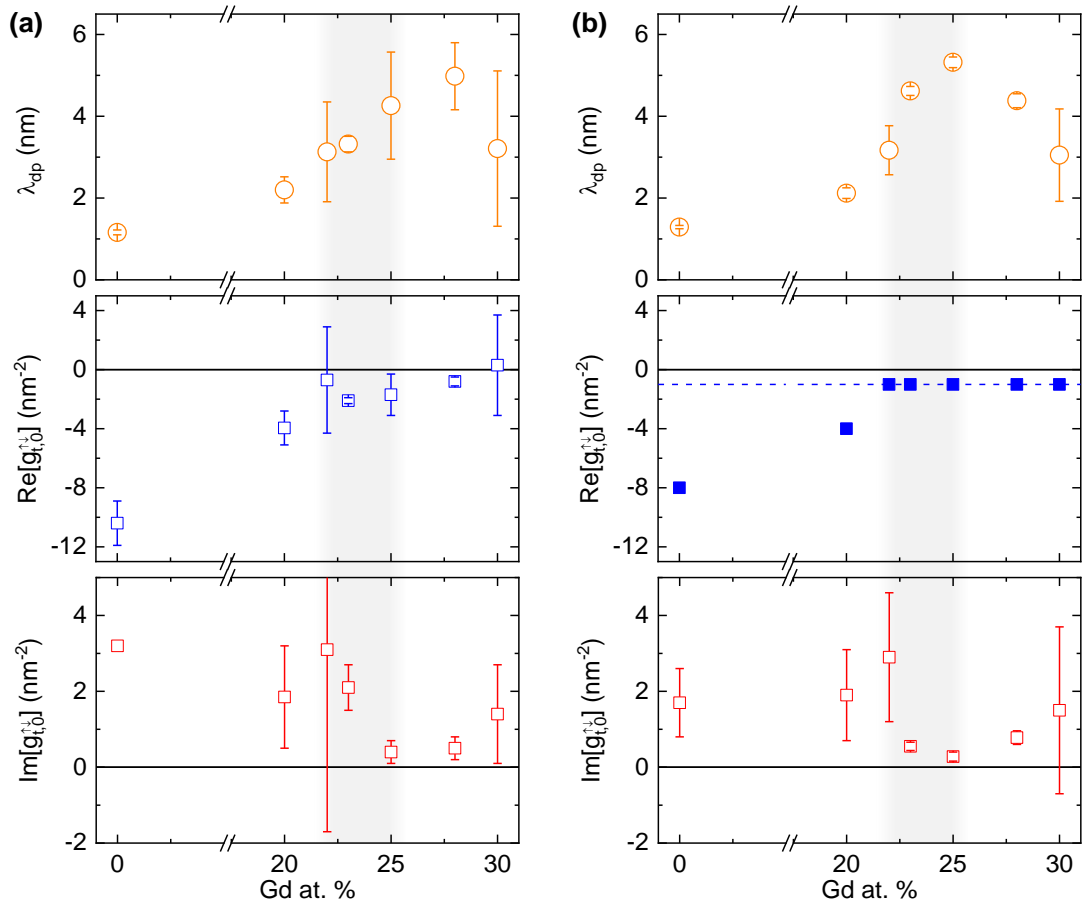

**FIG. 16.** Comparison of modeling results with (a) free $\mathrm{Re}[g_{t,0}^{\uparrow\downarrow}]$ and (b) fixed $\mathrm{Re}[g_{t,0}^{\uparrow\downarrow}]$. The shaded region indicates the window of composition corresponding to magnetic compensation.

**APPENDIX D. MODELING SPIN DEPHASING IN A FERRIMAGNETIC HETEROSTRUCTURE**

To understand the influence of the (collinear) magnetic order on the reflected and transmitted mixing conductances, we consider a magnetic trilayer composed of two equivalent nonmagnetic leads and a ferrimagnetic spacer, as illustrated in Fig. 17. We compute the mixing conductances of this system assuming coherent ballistic transport.

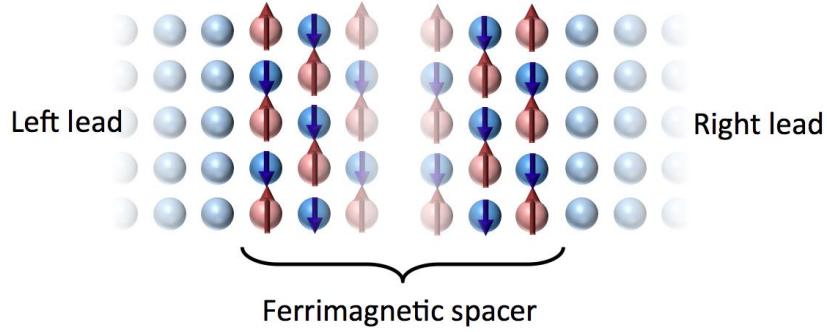

**FIG. 17.** Schematic of a ferrimagnetic trilayer as modeled below. The two magnetic sublattices are denoted by red and blue arrows pointing in opposite directions.

*System definition and boundary conditions:* Each layer is made of a three-dimensional square lattice with equal lattice parameter *a* for simplicity. The ferrimagnet is composed of a two-atomic unit cell with sublattices A and B. In the {A,B} basis, the Hamiltonian for spin $\sigma$ reads

$$\mathcal{H} = \begin{pmatrix} \varepsilon_A + \sigma \Delta_A - 4t_A \chi_k & -2t_{AB}\gamma_k \\ -2t_{AB}\gamma_k & \varepsilon_B + \sigma \Delta_B - 4t_B \chi_k \end{pmatrix}$$

with

$$\chi_k = \cos k_x a \cos k_y a + \cos k_x a \cos k_z a + \cos k_z a \cos k_y a$$

$$\gamma_k = \cos k_x a + \cos k_y a + \cos k_z a$$

In this expression, $\varepsilon_i$, $t_i$, $\Delta_i$ are the onsite energy, hopping integral and magnetic exchange on sublattice *i*, and $t_{AB}$ is the inter-sublattice hopping integral. The energy dispersion for spin $\sigma$ and band $\eta$ is

$$\varepsilon_{\eta,\sigma} = \bar{\varepsilon} + \sigma\bar{\Delta} - 4\bar{t}\chi_k + \eta\sqrt{(\delta\varepsilon + \sigma\delta\Delta - 4\delta t \chi_k)^2 + 4t_{AB}^2 \gamma_k^2}$$

and the associated eigenstate function reads

$$\hat{\phi}_{\eta,\sigma} = \frac{1}{\sqrt{2}}\sqrt{1 + \eta\beta_{k,\sigma}}|A\rangle \otimes |\sigma\rangle - \frac{\eta}{\sqrt{2}}\sqrt{1 - \eta\beta_{k,\sigma}}|B\rangle \otimes |\sigma\rangle$$

$$\beta_{k,\sigma} = \frac{\delta\varepsilon + \sigma\delta\Delta - 4\delta t \chi_k}{\sqrt{(\delta\varepsilon + \sigma\delta\Delta - 4\delta t \chi_k)^2 + 4t_{AB}^2 \gamma_k^2}}$$

In the above expressions, $\delta\varepsilon = \frac{\varepsilon_A - \varepsilon_B}{2}, \bar{\varepsilon} = \frac{\varepsilon_A + \varepsilon_B}{2}, \delta\Delta = \frac{\Delta_A - \Delta_B}{2}, \bar{\Delta} = \frac{\Delta_A + \Delta_B}{2}, \delta t = \frac{t_A - t_B}{2}, \bar{t} = \frac{t_A + t_B}{2}$.

Let us now build the scattering wave function across the trilayer. The electron wave functions in the left, central and right layers read

$$\psi_\sigma^L(\mathbf{k}) = \left(e^{i[k_x^L(\mathbf{k}_\perp)x + \mathbf{k}_\perp \cdot \boldsymbol{\rho}]} + \int \frac{d^2\boldsymbol{\kappa}}{4\pi^2} r_\sigma(\boldsymbol{\kappa}) e^{-i[k_x^L(\boldsymbol{\kappa})x + \boldsymbol{\kappa} \cdot \boldsymbol{\rho}]}\right)|C\rangle$$

$$\psi_\sigma^F(\mathbf{k}) = \int \frac{d^2\boldsymbol{\kappa}}{4\pi^2} \sum_\eta e^{i\boldsymbol{\kappa} \cdot \boldsymbol{\rho}}[A_{\eta,\sigma}\cos k_x^{\eta,\sigma}(\boldsymbol{\kappa})x + B_{\eta,\sigma}\cos k_x^{\eta,\sigma}(\boldsymbol{\kappa})x]\hat{\phi}_{\eta,\sigma}$$

$$\psi_\sigma^R(\mathbf{k}) = \int \frac{d^2\boldsymbol{\kappa}}{4\pi^2} t_\sigma(\boldsymbol{\kappa}) e^{i[k_x^R(\boldsymbol{\kappa})(x-d) + \boldsymbol{\kappa} \cdot \boldsymbol{\rho}]}|C\rangle$$

Here, $k_x^{L,R}(\mathbf{k}_\perp)$ is a solution of the dispersion relation in the leads, $\varepsilon(\mathbf{k}) = \varepsilon_N - 2t_N\gamma_k = \varepsilon_F$, $\varepsilon_F$ being the Fermi energy, and $\mathbf{k}_\perp$ is the in-plane component of the incoming wave vector. Similarly, $k_x^{\eta,\sigma}(\boldsymbol{\kappa})$ is determined by the condition $\varepsilon_{\eta,\sigma}(\mathbf{k}) = \varepsilon_F$. Now, let us determine the matching conditions at $x = 0$ and $x = d$. Because the leads and the ferrimagnetic layer possess a different first Brillouin zone, there will be Umklapp scattering [90] that must be taken into account in the matching procedure. In fact, the first Brillouin zone of the ferrimagnet is twice smaller than the first Brillouin zone of the leads, introducing an Umklapp momentum $\mathbf{Q} = \frac{\pi}{a}(\mathbf{y} + \mathbf{z})$ [90–92]. We then project the wave function of the ferrimagnetic layer on the normal metal orbital $|C\rangle$ and use the fact that $\langle C|B\rangle = \langle C|A\rangle e^{i(\kappa_y - k_{\perp,y})a}$. In summary, the boundary conditions for normal scattering are

$$1 + r_\sigma = \sum_\eta \left[\frac{\eta}{\sqrt{2}}\sqrt{1 - \eta\sigma\beta} + \frac{1}{\sqrt{2}}\sqrt{1 + \eta\sigma\beta}\right] A_{\eta,\sigma}$$

$$ik_x^L(1 - r_\sigma) = \sum_\eta \left[\frac{\eta}{\sqrt{2}}\sqrt{1 - \eta\sigma\beta} + \frac{1}{\sqrt{2}}\sqrt{1 + \eta\sigma\beta}\right] k_x^\eta B_{\eta,\sigma}$$

$$\sum_\eta \left[\frac{\eta}{\sqrt{2}}\sqrt{1 - \eta\sigma\beta} + \frac{1}{\sqrt{2}}\sqrt{1 + \eta\sigma\beta}\right](A_{\eta,\sigma}\cos k_x^\eta d + B_{\eta,\sigma}\sin k_x^\eta d) = t_\sigma$$

$$\sum_\eta \left[\frac{\eta}{\sqrt{2}}\sqrt{1 - \eta\sigma\beta} + \frac{1}{\sqrt{2}}\sqrt{1 + \eta\sigma\beta}\right] k_x^\eta (-A_{\eta,\sigma}\sin k_x^\eta d + B_{\eta,\sigma}\sin k_x^\eta d) = ik_x^R t_\sigma$$

and for the Umklapp scattering, we obtain

$$d_\sigma = \sum_\eta \left[\frac{\eta}{\sqrt{2}}\sqrt{1 - \eta\sigma\beta_Q} - \frac{1}{\sqrt{2}}\sqrt{1 + \eta\sigma\beta_Q}\right] A_{\eta,\sigma}$$

$$-ik_{x,Q}^L d_\sigma = \sum_\eta \left[\frac{\eta}{\sqrt{2}}\sqrt{1 - \eta\sigma\beta_Q} - \frac{1}{\sqrt{2}}\sqrt{1 + \eta\sigma\beta_Q}\right] k_{x,Q}^\eta B_{\eta,\sigma}$$

$$\sum_{\eta} \left[ \frac{\eta}{\sqrt{2}} \sqrt{1 - \eta\sigma\beta_Q} - \frac{1}{\sqrt{2}} \sqrt{1 + \eta\sigma\beta_Q} \right] (A_{\eta,\sigma} \cos k_{x,Q}^{\eta} d + B_{\eta,\sigma} \sin k_{x,Q}^{\eta} d) = u_{\sigma}$$

$$\sum_{\eta} \left[ \frac{\eta}{\sqrt{2}} \sqrt{1 - \eta\sigma\beta_Q} - \frac{1}{\sqrt{2}} \sqrt{1 + \eta\sigma\beta_Q} \right] k_{x,Q}^{\eta} (-A_{\eta,\sigma} \sin k_{x,Q}^{\eta} d + B_{\eta,\sigma} \sin k_{x,Q}^{\eta} d) = i k_{x,Q}^{R} u_{\sigma}$$

In order to make the above expressions easier to track, we have set

$$r_{\sigma} = r_{\sigma}(\mathbf{k}_{\perp}), t_{\sigma} = t_{\sigma}(\mathbf{k}_{\perp}), \beta = \beta_{\mathbf{k}_{\perp}}, k_x^{L,R} = k_x^{L,R}(\mathbf{k}_{\perp}), k_x^{\eta} = k_x^{\eta}(\mathbf{k}_{\perp})$$

$$d_{\sigma} = r_{\sigma}(\mathbf{k}_{\perp} + \mathbf{Q}), u_{\sigma} = t_{\sigma}(\mathbf{k}_{\perp} + \mathbf{Q}), \beta_Q = \beta_{\mathbf{k}_{\perp}+\mathbf{Q}}, k_{x,Q}^{L,R} = k_x^{L,R}(\mathbf{k}_{\perp} + \mathbf{Q}), k_{x,Q}^{\eta} = k_x^{\eta}(\mathbf{k}_{\perp} + \mathbf{Q})$$

*Mixing Conductances:* The reflected and transmitted mixing conductances are defined [93]

$$g_r^{\uparrow\downarrow} = \left( \frac{e^2}{h} \right) \int \frac{d^2\boldsymbol{\kappa}}{4\pi^2} (1 - r_{\uparrow} r_{\downarrow}^* - d_{\uparrow} d_{\downarrow}^*)$$

$$g_t^{\uparrow\downarrow} = \left( \frac{e^2}{h} \right) \int \frac{d^2\boldsymbol{\kappa}}{4\pi^2} (t_{\uparrow} t_{\downarrow}^* + u_{\uparrow} u_{\downarrow}^*)$$

We now compute these two quantities (both real and imaginary parts) as a function of the ferrimagnetic layer thickness upon varying the magnetic exchange. For these calculations, we set $\varepsilon_N = \bar{\varepsilon} = 2.5$ eV, $\delta\varepsilon = 0, t_N = t_{AB} = 1$ eV, $\bar{t} = 0$. We also set $\Delta_A = 1$ eV and $\Delta_B$ is varied between +1 eV (ferromagnet) and -1 eV (antiferromagnet). The results are reported in Fig. 18. The reflected mixing conductance is weakly affected by the magnetic order, which is expected based on simple free-electron arguments [93], whereas the transmitted mixing conductance is dramatically modified when tuning the magnetic exchange. In the ferromagnetic limit ($\Delta_B = +1$ eV), it displays the expected damped oscillatory behavior already observed [93] and attributed to the destructive interferences between precessing spins with different incidence (in other words, spin dephasing). Reducing and then inverting the exchange leads to an overall reduction of the average exchange field, leading to an increase of the dephasing length, which diverges in the antiferromagnetic limit ($\Delta_B = -1$ eV). We notice that the imaginary part of the transmitted mixing conductance, $\text{Im}[g_t^{\uparrow\downarrow}]$, vanishes in the antiferromagnetic limit, qualitatively consistent with our experimental observation.

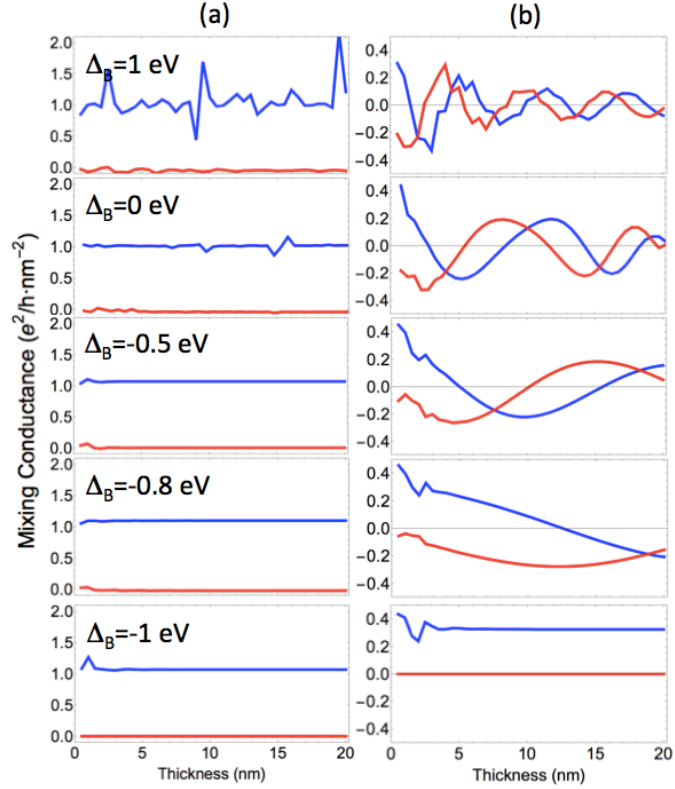

**FIG. 18.** Dependence of the (a) reflected and (b) transmitted mixing conductance as a function of the ferrimagnetic layer thickness upon varying the exchange on sublattice B. The real (imaginary) part of the mixing conductance is reported in blue (red). The unit is in ($e^2/h$) and per nm².

For the sake of completeness, the (inverse of the) dephasing length $\lambda_{dp}$ is reported in Fig. 19 as a function of the magnetic exchange of the sublattice B. One clearly sees a mostly linear dependence that suggests $\lambda_{dp} \sim 1/(\Delta_A + \Delta_B)$.

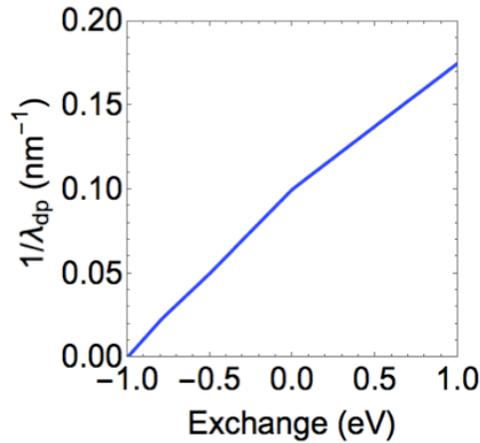

**FIG. 19.** Inverse of the dephasing length as a function of the magnetic exchange.